\documentclass[pra,preprint,aps,superscriptaddress,showpacs]{revtex4}
\usepackage{amsmath,amssymb,latexsym,amsfonts}
\usepackage{epsfig}
\begin{document}
\title{Determination of Compton profiles at solid surfaces from first-principles calculations}
\date{\today}

\author{Christoph Lemell}
\email[Author, correspondence should be sent to; Email: ]{lemell@concord.itp.tuwien.ac.at}
\affiliation{Institute for Theoretical Physics, Vienna University of 
Technology, Wiedner Hauptstra\ss e 8-10, A--1040 Vienna, Austria}
\author{Andres Arnau}
\affiliation{Departamento de Fisica de Materiales UPV/EHU and Centro Mixto C.S.I.C.-UPV, San Sebastian, Spain, EU}
\author{Joachim Burgd\"orfer}
\affiliation{Institute for Theoretical Physics, Vienna University of 
Technology, Wiedner Hauptstra\ss e 8-10, A--1040 Vienna, Austria}

\begin{abstract}
Projected momentum distributions of electrons, i.e.\ Compton profiles above the topmost atomic layer have recently become experimentally accessible by kinetic electron emission in grazing-incidence scattering of atoms at atomically flat single crystal metal surfaces. Sub-threshold emission by slow projectiles was shown to be sensitive to high-momentum components of the local Compton profile near the surface. We present a method to extract momentum distribution, Compton profiles, and Wigner and Husimi phase space distributions from ab-initio density-functional calculations of electronic structure. An application for such distributions to scattering experiments is discussed.
\end{abstract}

\maketitle

\section{Introduction}
\label{intro}

The local density of states above the topmost layer, the occupied states of which represent the electron density spilling out into vacuum, is an important quantity characterizing many properties of the surface and image states \cite{jb1,jb2}, surface magnetism \cite{jb3,jb3a}, and chemical reactivity mediated by charge transfer \cite{jb4,jb5}. Yet, many of its properties are difficult to directly access experimentally because they are overshadowed by bulk contributions. A case in point is the projected momentum distribution, the Compton profile of the electronic structure. While bulk Compton profiles have been mapped out in considerable detail by electron scattering \cite{jb6}, fast ion scattering \cite{jb6a}, and X-ray scattering \cite{jb7}, little is known about the surface Compton profile characterizing the anisotropic momentum distribution of the ground-state electronic structure reaching out into vacuum.

Recent experiments for grazing-incidence scattering of low energy neutral rare-gas atoms at Al(111) and Al(110) single crystal surfaces provide first evidence of high-momentum components in the electronic selvage \cite{prb}. The projectile threshold velocity $v_{th}$ required for the kinetic emission of surface electrons by atoms in a head-on binary encounter collision $(2v_{th}+v_e)^2/2\geq (E_F+W)$ ($E_F$: Fermi energy, $W$: work function) was found to be below the standard value $v_{th}^0$ \cite{bara} for $v_e=v_F$ ($v_F$: Fermi velocity). This sub-threshold kinetic emission indicates the presence of off-shell velocity (momentum) components well above the Fermi velocity $v_e=v_F$. Obviously, the frequently employed free-electron gas approximation for conduction band electrons is not sufficiently accurate to account for sub-threshold kinetic electron emission (KE) \cite{prb}. The experimental data can be taken as first evidence for the presence of momenta above the Fermi momentum $q_F$ in realistic momentum distributions above the surface taking many-particle effects (correlation) and the crystal potential into account.

In the present communication we discuss the extraction of the momentum distribution and related quantities from ground-state density-functional calculations of the surface band structure. For estimating impact-parameter dependent emission probabilities in grazing-incidence surface scattering information beyond the projected momentum distribution, the Compton profile, is desirable: the quantum phase-space distribution or Wigner function \cite{jb10}, $\rho^W(\vec q,\vec r)$. It allows to determine the correlation between position above the surface and local momentum encoded in the electron density compatible with the Heisenberg uncertainty principle. We also consider the Husimi distribution \cite{jb11}, $\rho^H(\vec q,\vec r)$, which represents a minimum-wavepacket average over $\rho^W$.

We present numerical results for the Al(111) and Al(110) single crystal surfaces using the ABINIT code \cite{abinit} with soft pseudopotentials \cite{rappe}. Applications to KE will be discussed. Atomic units are used throughout unless stated otherwise.

\section{Momentum distributions and density functional theory}
State-of-the-art electronic structure calculations for the bulk and surface invoke density-functional theory (DFT \cite{jb12}, for electronic structure programs employing DFT see, e.g.\ \cite{abinit,vasp,siesta,wien2k}). Ground-state energies as well as excitation spectra can be calculated within linear-response with sub-eV accuracy. Apart from the ubiquitous uncertainty with regard to the appropriate exchange-correlation potential, $V_{xc}$, application of DFT to other classes of observables faces the conceptual difficulty that appropriate read-out functionals are, a priori, not known unless the observable can be directly expressed in terms of the electronic density $\rho(\vec r)$. In the present contribution we inquire into a read-out functional for the Compton profile, i.e.\ the momentum distribution $\rho(\vec q)$ near surfaces.

DFT employs the solution of a set of one-particle Kohn-Sham (KS) equations for the pseudo-wavefunctions $\psi_l$ \cite{ks}
\begin{equation}
\label{eq:2}
H_{eff}(\vec r)\psi_l(\vec r)=\left(-\frac{1}{2}\Delta+V_{eff}[\rho(\vec r)]\right)\psi_l(\vec r)=\varepsilon_l\psi_l(\vec r)
\end{equation}
which have the property to add up to the exact electron density
\begin{equation}
\rho(\vec r)=\sum_l |\psi_l(\vec r)|^2\, ,
\label{eq:3}
\end{equation}
provided the exact functional form of the exchange-correlation potential $V_{xc}$ entering the effective potential $V_{eff}$ is known. This, however, is not the case and a variety of approximate functionals suited for specific problems are in use.

The quasicontinuum of bands $n$ filled up to a maximum energy $E_F$ is replaced by a discretized set of wavefunctions with band indices $n$ and $\vec k$. Then, for any band with band index $n$ the wavefunction associated with momentum $\vec k$ in the first Brioullin zone solving the Schr\"odinger equation for the solid can be written by a sum over plane waves
\begin{equation}
\psi_{n,\vec k}(\vec r)=\sum_{\vec G}c_{n,\vec k}(\vec G)\, 
e^{i(\vec k+\vec G)\vec r}
\label{eq:4}
\end{equation}
where $\vec G$ is any reciprocal lattice vector. Employing periodic boundary conditions, first-principle codes compute the electron density either for infinitely extended crystals (computational box is equal to the crystal unit cell) or, at crystal surfaces, by calculating the ground state for an infinite number of thin crystal slabs (``supercell methods''). In either case, the coefficients $c_{n,\vec k}$ are returned as output of the calculation. The total electron density is given by,
\begin{equation}
\rho(\vec r)=\sum_n \sum_{\vec k} \mbox{occ}_{n,\vec k} w_{\vec k} \left|\psi_{n,\vec k}(\vec r)\right|^2
\label{eq:5}
\end{equation}
where $\mbox{occ}_{n,\vec k}$ is the occupation of the (pseudo-) state with indices $n$ and $\vec k$ and $w_{\vec k}$ is the weight for the point $\vec k$ in reciprocal space resulting from symmetry considerations if calculations are restricted to the irreducible Brioullin zone. In our case, $w_{\vec k}=const$ as we work on an equal-spaced grid extended over the whole Brioullin zone.

\subsection{Momentum distributions $\rho(\vec q)$}
\label{2.1}

For a Schr\"odinger wavefunction $\phi(\vec r)$ coordinate-space densities and momentum densities are related through the Fourier transform of the wavefunction,
\begin{equation}
\rho(\vec q)=\left | \tilde\phi (\vec q)\right|^2
\label{eq_jb1}
\end{equation}
with
\begin{equation}
\tilde\phi (\vec q)=(2\pi)^{-3/2}\int d^3r\, e^{-i\vec q\vec r}\,\phi(\vec r)\, .
\label{eq_jb2}
\end{equation}
Since the KS pseudo-wavefunctions $\psi_{n,\vec k}(\vec r)$ are, in general, not to be identified with true wavefunctions, the applicability of Eqs.\ \ref{eq_jb1} and \ref{eq_jb2} is, a priori, not obvious. Note that $\rho(\vec r)$ and $\rho(\vec q)$ are not Fourier transforms of each other. As an alternative to applying Eq.\ \ref{eq_jb2} to KS-wavefunctions, it might be possible to formulate the Hohenberg-Kohn theorem in momentum space with $\rho(\vec q)$ as fundamental quantity. In this case the KS equations would become integral equations and $V_{xc}$ a non-local integral operator. To our knowledge, this avenue has not yet been explored.

In the following, we postulate the applicability of Eqs.\ \ref{eq_jb1} and \ref{eq_jb2} for KS pseudo-wavefunctions. This can be justified for the homogeneous electron gas. Its extension to realistic DFT calculations for surfaces can be viewed as an analogue to the local-density approximation to the exchange-correlation potential.

\begin{figure}
\centerline{\epsfig{file=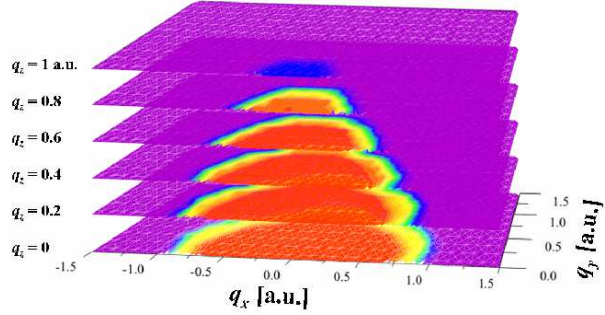,width=8cm}}
\caption{(Color online) Momentum distribution of an Al(111) slab. The Fermi momentum for Al is about $q_F\approx 0.9$ a.u. Higher momenta represent correlation and crystal structure effects.}
\label{momdist}
\end{figure}

Accordingly, $\psi_{n,\vec k}$ is given in momentum representation by
\begin{eqnarray}
\tilde\psi_{n,\vec k}(\vec q)&=&(2\pi)^{-3/2}\int \sum_{\vec G}c_{n,\vec k}(\vec G)\, 
e^{i(\vec k+\vec G)\vec r}\, e^{-i\vec q\vec r}\, d^3r\nonumber\\
&=&(2\pi)^{-3/2}\sum_{\vec G}c_{n,\vec k}(\vec G)\,\delta(\vec q-(\vec k+\vec G))\, .
\label{eq:6}
\end{eqnarray}
For the three-dimensional momentum distribution we find
\begin{eqnarray}
\rho(\vec q)&=&\sum_{n,\vec k}\mbox{occ}_{n,\vec k}\,w_{\vec k}\,
\left | \tilde\psi_{n,\vec k}(\vec q) \right|^2\nonumber\\
&=&(2\pi)^{-3}\sum_{n,\vec k}\mbox{occ}_{n,\vec k}\,w_{\vec k}\,
\sum_{\vec G,\vec G'}c_{n,\vec k}(\vec G)c_{n,\vec k}^*(\vec G')\,
\delta(\vec q-(\vec k+\vec G))\delta(\vec q-(\vec k+\vec G'))\nonumber\\
&=&(2\pi)^{-3}\sum_{n,\vec k}\mbox{occ}_{n,\vec k}\,w_{\vec k}\,
\sum_{\vec G}\left | c_{n,\vec k}(\vec G)\right |^2 \delta(\vec q-(\vec k+\vec G))\, .
\label{eq:7}
\end{eqnarray}
Fig.\ \ref{momdist} shows cuts through the momentum distribution of an Al(111) single crystal slab at fixed $q_z$ values. As Al is well described by a nearly-free electron approximation, the momentum distribution closely resembles a sphere with radius $q_F\approx 0.9$ a.u.\ but features also small but finite contributions at higher momenta due to correlation and the crystal structure potential. Furthermore, calculating $\rho(\vec q)$ for a crystal slab leads to an additional increase of higher momenta normal to the surface ($q_z$) due to the breaking of translational symmetry at the surfaces.

\subsection{Distance dependent momentum distributions $\rho(\vec q_\|,z)$}
\label{2.2}

\begin{figure}
\centerline{\epsfig{file=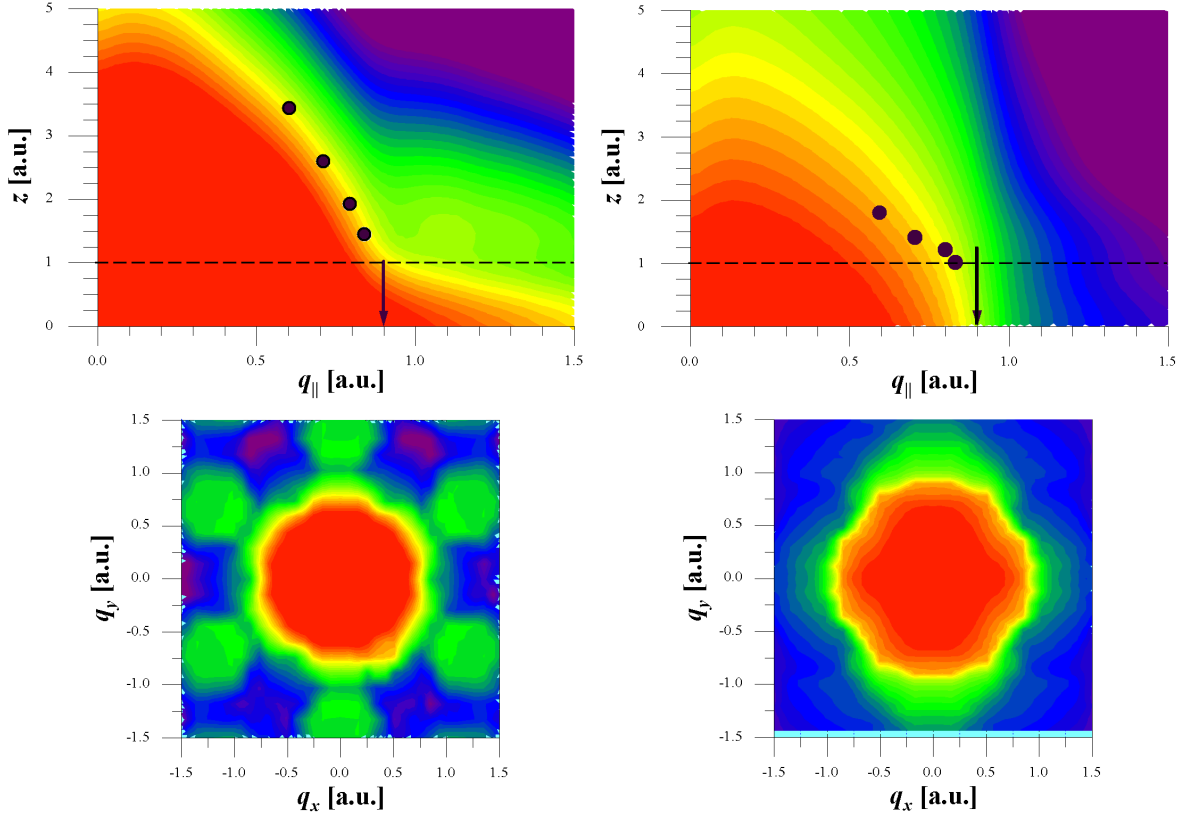,width=16cm}}
\caption{(Color online) $z$-dependent momentum distributions as a function of the distance from the topmost atomic layer. Top left and right panels show distributions for Al(111) and Al(110) surfaces, respectively. Black dots indicate experimental estimates for local Fermi momenta (see chapter \ref{app}). Bottom panels show $q_x - q_y$ distributions for both faces taken at $z=1$ a.u.\ above the topmost atomic layer (dashed line in top panels). The white arrow in the bottom left panel indicates the wavenumber $q=2\pi/a_s\approx 1.16$ related to the nearest-neighbor distance of Al(111) surfaces.}
\label{zdep}
\end{figure}

If we perform the Fourier transform only for coordinates parallel to the surface we retain the $z$-dependence (coordinate along the surface normal) of the position distribution but extract the momentum distribution along the coordinates in the surface plane $\rho(\vec q_\|,z)$. It can be rewritten as a sum over plane-wave coefficients and delta functions,
\begin{eqnarray}
\rho(\vec q_\|,z)&=&\sum_{n,\vec k}\mbox{occ}_{n,\vec k}\,w_{\vec k}\,
\left | \tilde\psi_{n,\vec k}(\vec q_\|,z)\right |^2\nonumber \\
&=&(2\pi)^{-2}\sum_{n,\vec k}\mbox{occ}_{n,\vec k}\,w_{\vec k}\,
\sum_{\vec G,\vec G'} c_{n,\vec k}(\vec G)c_{n,\vec k}^*(\vec G')\,
e^{i(G_z-G'_z)z}\times\\
&&\qquad\qquad\qquad\qquad\qquad\times\; \delta(\vec q_\|-(\vec k_\|+\vec G_\|))
\delta(\vec q_\|-(\vec k_\|+\vec G'_\|))\, .\nonumber
\label{eq:8}
\end{eqnarray}
Fig.\ \ref{zdep} shows the distance dependent $|q_\| |$-distributions for Al(111) (left panels) and
Al(110) surfaces (right panel) averaged over the azimuthal angle $\varphi_q$ as a function of distance $z$ from the topmost atomic layer on a logarithmic scale (color coding). The Fermi momentum is indicated by the black arrows. Differences between the distributions provide information on the face dependence of the momentum distributions and on the underlying surface potential. For example, the hump near the wavenumber $q=2\pi/a_s\approx 1.16$ a.u.\ in the top left panel of Fig.\ \ref{zdep} originates from the nearest neighbor distance in the Al(111) surface $a_s\approx 5.4$ a.u. In the case of the Al(110) surface (right hand panels) the existence of ``wide'' and ``narrow'' surface channels becomes evident by the different extension of the distribution in $q_x$ and $q_y$ directions (bottom right).

\subsection{Wigner distributions $\rho^W(\vec q,\vec r)$}
\label{2.3}

The Wigner distribution is the quantum mechanical analogue of the classical phase-space distribution $\rho^{cl}(\vec q,\vec r)$. It features remarkable resemblance to $\rho^{cl}(\vec q,\vec r)$ as it fulfills the same relations as $\rho^{cl}$
\begin{eqnarray}
\rho(\vec q)&=&\int d^3\vec r\, \rho^W(\vec q,\vec r)\label{eq:9b}\\
\rho(\vec r)&=&\int d^3\vec q\, \rho^W(\vec q,\vec r)\, .\label{eq:9c}
\end{eqnarray}
$\rho^W$ is real valued. An interpretation in terms of a probability distribution is, however, not possible as $\rho^W$ is not positive definite. The latter is the immediate consequence of the position-momentum uncertainty. $\rho^W$ can be calculated from the Fourier transform of the off-diagonal elements of the single-particle density matrix
\begin{equation}
\hat\rho=\sum_{n,\vec k}\mbox{occ}_{n,\vec k}\,w_{\vec k}\,|\psi_{n,\vec k}\rangle
\langle\psi_{n,\vec k}|\, ,
\label{eq:9}
\end{equation}
and is given by
\begin{equation}
\rho^W(\vec q,\vec r)=\frac{1}{\pi^3}\int
\langle \vec r-\vec y|\hat \rho |\vec r +\vec y\rangle \,
e^{2i\vec q\vec y}\, d^3y\, .
\label{eq:10}
\end{equation}

\begin{figure}
\centerline{\epsfig{file=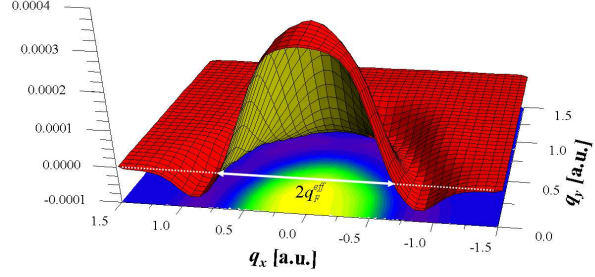,width=8cm}}
\caption{(Color online) Wigner function $\rho^W(\vec q,\vec r)$ in front of an Al(111) surface. $q_z=0,\, z=1$ a.u.\ above a surface atom. Around the main part of the distribution, negative valued areas (purple contour, also in Fig.\ \ref{wig2}) impede the interpretation as probability distribution.}
\label{wig1}
\end{figure}

Assuming now again that the KS orbitals can be taken to represent the one-particle density matrix, the calculation of $\rho^W$ using DFT input is straightforward. It should be noted that the use of KS-orbitals assures that the exact position density results from the reduction of the Wigner function (Eq.\ \ref{eq:9c}). This observation suggests that the representation of $\rho^W$ in terms of KS orbitals and thus also of $\rho(\vec q)$ (Eq.\ \ref{eq:9b}) is a meaningful approximation. For a single-band wavefunction $\psi_{n,\vec k}$ we find
\begin{eqnarray}
\rho^W_{n,\vec k}(\vec q,\vec r)&\propto&\int\sum_{\vec G,\vec G'}
c_{n,\vec k}(\vec G)e^{i(\vec k+\vec G)(\vec r-\vec y)}
c_{n,\vec k}^*(\vec G')e^{-i(\vec k+\vec G')(\vec r+\vec y)}
e^{2i\vec q\vec y}\, d^3y\nonumber\\
&=&\sum_{\vec G,\vec G'}c_{n,\vec k}(\vec G)
c_{n,\vec k}^*(\vec G')e^{i(\vec G-\vec G')\vec r}
\int e^{iy(2\vec q-2\vec k-\vec G-\vec G')}\, d^3y\nonumber\\
&\propto&\sum_{\vec G,\vec G'}c_{n,\vec k}(\vec G)
c_{n,\vec k}^*(\vec G')e^{i(\vec G-\vec G')\vec r}\,
\delta (2\vec q-2\vec k-\vec G-\vec G')\, .\label{eq:12a}
\end{eqnarray}
The total Wigner function is therefore
\begin{equation}
\rho^W(\vec q,\vec r)\propto
\sum_{n,\vec k}\mbox{occ}_{n,\vec k}\,w_{\vec k}\;\rho^W_{n,\vec k}(\vec q,\vec r)\, .
\label{eq:12b}
\end{equation}
The failure of the naive interpretation as a probability distribution can be seen from Fig.\ \ref{wig1}: $\rho^W$ always features negative regions and strong oscillations in the regions of interest.
\begin{figure}
\centerline{\epsfig{file=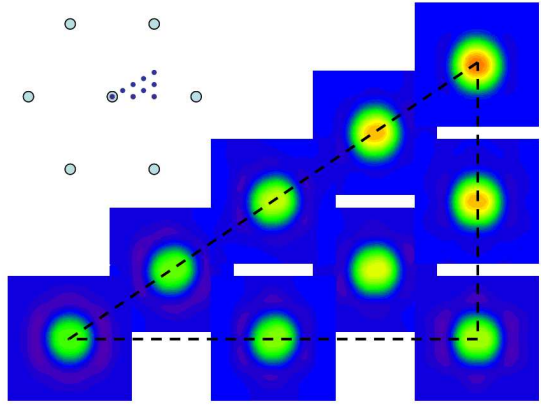,width=8cm}}
\caption{(Color online) Wigner function for an Al(111) surface evaluated at 9 points of the irreducible part of the surface unit cell (see inset) and $q_z=0\, z=1$ a.u.}
\label{wig2}
\end{figure}
Fig.\ \ref{wig2} shows cuts through the Wigner function for various positions above the irreducible part of the surface unit cell (see inset) at $q_z=0$ a.u. As Wigner functions are a natural starting point for quantum calculations of scattering processes, details of the distributions may account for features in electron emission experiments sensitive to details of the projectile trajectory along the surface.

\subsection{Husimi distributions $\rho^H(\vec q,\vec r)$}
\label{2.4}

The Husimi distribution is a convoluted (or averaged) Wigner function and even closer linked to classical phase-space distribution than the Wigner function. Husimi suggested to average the Wigner function over a minimum uncertainty wavepacket. $\rho^H$ is defined by
\begin{equation}
\rho^H(\vec q,\vec r)=\frac{1}{\pi^3}\int d^3 \vec x \int d^3\vec p\,
\rho^W(\vec p,\vec x)\exp\left[-\frac{(\vec p-\vec q)^2}{\sigma^2}-\sigma^2(\vec x-\vec r)^2\right]
\label{eq:13}
\end{equation}
where $\sigma$ determines the width of the wavepacket in $\vec q$ and $\vec r$ directions. $\rho^H$ is the closest analogue to classical phase-space distribution consistent with the Heisenberg uncertainty principle. In particular, $\rho^H$ is positive definite and allows for a probability interpretation. Eq.\ \ref{eq:13} would be time consuming to evaluate directly as it would require the evaluation of $\rho^W$ for all coordinates $\vec q$ and $\vec r$ and performing a 6-dimensional integration. This, however, can be circumvented by resorting to the equivalent expressions
\begin{equation}
\rho_{nk}^H (\vec q,\vec r) \propto \left| {\int {\psi_{n,\vec k} (\vec x)\exp \left\{ { - \frac{1}{{2\sigma^2}}(\vec x-\vec r)^2  + i\vec q\vec x} \right\}d^3 \vec x} } \right|^2 
\label{eq:14}
\end{equation}
and
\begin{equation}
\rho^H(\vec q,\vec r)\propto
\sum_{n,\vec k}\mbox{occ}_{n,\vec k}\,w_{\vec k}\;\rho^H_{n,\vec k}(\vec p,\vec r)\, ,
\label{eq:15}
\end{equation}
which reduces the numerical effort considerably. $\sigma$ in Eq.\ \ref{eq:14} again determines the width of the wavepacket in $\vec r$. The averaging process smooths out oscillations of $\rho^W$ on the scale of the de Broglie wavelength. It renders $\rho^H(\vec q,\vec r)$ positive definite over the whole phase space and therefore allows for the desired interpretation as the probability of finding an electron with momentum $\vec q$ at position $\vec r$. It has, however, the disadvantage of featuring unphysically high momentum components due to the infinitely long tail of the Gaussian distribution in Eqs.\ \ref{eq:13} and \ref{eq:14}. Employing such a distribution in scattering calculations will therefore lead to a spontaneous escape of electrons with energies higher than the surface potential unless a suitable projection formalism is applied. Alternatively, reducing the width of the wavepacket in momentum space would also diminish the weight of the high-momentum components but would come at the price of an averaging over a large volume in $\vec r$ space (and therefore $z$) smoothing out the density gradient at the surface. The Husimi distribution can therefore be mainly used for an intuitive interpretation of quantum mechanical results in terms of the quasi-classical probability distribution.

\section{Application to above surface kinetic electron emission}
\label{app}
Outside the topmost atomic layer the electron density is quickly reduced. Electron wavefunctions reaching farthest out of the surface will have large momentum in direction normal to the surface ($q_z$). Therefore, smaller momenta parallel to the surface are observed than in the bulk. This behavior has recently been experimentally investigated by Winter et al.\ \cite{epl}. They have scattered atomic projectiles off single crystal Al(111), Al(110), and Cu(111) surfaces under surface channeling conditions. The turning point of the trajectory was varied by changing the angle of incidence relative to the surface normal. As expected, they find a decreasing maximum $q_\|$, referred to as ``local Fermi momentum'', with increasing distance from the surface. Their data can be directly compared to our results in sections \ref{2.2} and \ref{2.3}. We have displayed the experimental results in the top panels of Fig.\ \ref{zdep} for both Al surfaces. The contour levels were chosen as to approximately match the experimental data point for closest approach to the surface. In the case of Al(111) the agreement between theory and experiment is almost perfect whereas results for the Al(110) surface agree only on a qualitative level. This may have two reasons: on the one hand, trajectories along Al(111) surfaces are better defined than along Al(110). The Al(111) surface features the ``smoother'' planar surface potential due to its closest packed structure. Additionally, electron spill-out and therefore the experimental signal is larger in front of Al(111) surfaces than in front of Al(110) which may lead to a larger error in the experimentally determined local Fermi momentum.

\begin{figure}
\centerline{\epsfig{file=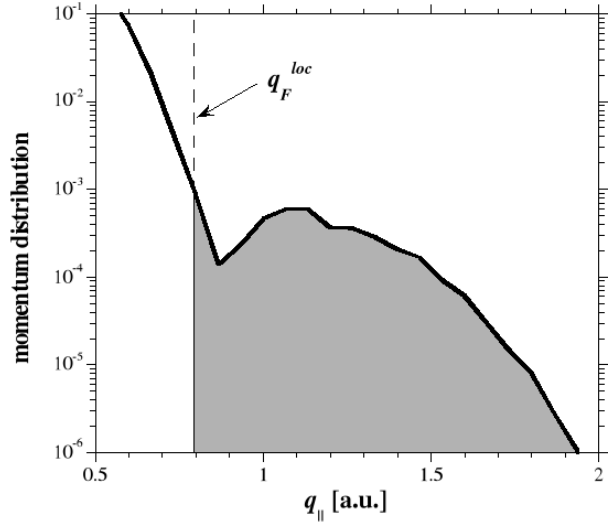,width=8cm}}
\caption{High-momentum tail of $\rho(q_\|,z)$ for $z=2$ a.u. The local Fermi momentum \cite{epl} is indicated by the dashed line. Contributions above $\rho_F^{loc}$ (shaded area) lead to sub-threshold kinetic electron emission.}
\label{JB_fig}
\end{figure}

As the determination of the local Fermi momentum is closely related to measurements of ``sub-threshold'' KE \cite{prb}. $z$-dependent momentum distributions are also used for its interpretation. In this case, momenta above the local Fermi momentum are related to the KE yield (shaded area in Fig.\ \ref{JB_fig}). KE for projectile velocities below $v_{th}^0$ (see Sec.\ \ref{intro}) cannot be explained within the non-interacting free electron model for metallic conduction bands.

\section{Conclusion}
We have proposed a method to extract information on the surface momentum distributions and Compton profiles from ab-initio DFT calculations. Pseudo-wavefunctions from such calculations are (partially) Fourier transformed in order to derive the momentum distributions $\rho(\vec q)$ and $\rho(\vec q_\|,z)$ and the quantum mechanical phase-space distributions $\rho^W(\vec q,\vec r)$ (Wigner distribution) and $\rho^H(\vec q,\vec r)$ (Husimi distribution). Far from resembling a simple Fermi distribution, we find momentum components well above $q_F$ originating from correlation effects and the periodic crystal potential of the solid. Two examples for applications of the calculated distributions are presented and compared to experimental results \cite{epl}.

\begin{acknowledgments}
The work was supported by the Austrian Fonds zur F\"orderung der wissenschaftlichen Forschung (FWF Austria) within the special research project SFB-016, EU contracts no.\ HPRI-CT-2001-50036 and no.\ HPRI-CT-2005-026015, and the Donostia International Physics Center. A.A. acknowledges support by UPV/EHU grant No.\ 9/UPV 00206.215-13639/2001, MEC grant No.\ FIS2004-06490-C03-00, and EU Network of excellence NANOQUANTA grant No.\ NMP4-CT-2004-500198.
\end{acknowledgments}

\end{document}